%% file: NYU_InterDigital_arXiv.tex
\documentclass[conference]{IEEEtran}
\IEEEoverridecommandlockouts

\usepackage{amsmath}
\usepackage{amsfonts}
\usepackage{amssymb}
\usepackage{graphicx}
\graphicspath{ {figures/} }
\usepackage{epsfig}
\usepackage{subfig}
\usepackage{float}
\usepackage{epstopdf}
\usepackage{array}
\usepackage{multirow}
\usepackage{setspace}
\usepackage[usenames,dvipsnames,table]{xcolor}
\usepackage{color}
\usepackage{fancyhdr}
\usepackage[section]{placeins}
\usepackage{url}
\usepackage{algorithm}
\usepackage{algpseudocode}
\usepackage{multicol}
\usepackage{lipsum}
\usepackage{soul}
\usepackage{comment}
\usepackage{balance}
\usepackage{xcolor}
\usepackage{cite}
\usepackage[normalem]{ulem}
\usepackage[acronym]{glossaries}
\usepackage{colortbl}
\input{acronyms} 

\usepackage{authblk}
\usepackage[top    = 0.7in,
bottom = 0.8in,
left   = 0.625in,
right  = 0.625in]{geometry}

\definecolor{TableColor}{RGB}{217,217,217}
\fancypagestyle{firststyle}{
	\fancyhf{}
	\fancyhead[L]{O.Kanhere, S. Goyal, M. Beluri, and T. S. Rappaport, ``Target Localization using Bistatic and Multistatic Radar with 5G NR Waveform,''  2021 \textit{IEEE 93rd Vehicular Technology Conference (VTC-Spring)}, April 2021, pp. 1-7.
	}                           

}
\IEEEoverridecommandlockouts
\begin{document}
	\title{Target Localization using Bistatic and Multistatic Radar with 5G NR Waveform \vspace{-0.2cm}}
	\author[1,2]{Ojas Kanhere\thanks{Portions of this work was supported by the NYU WIRELESS Industrial Affiliates Program, InterDigital Communications, Inc., and National Science Foundation (NSF) Research Grants: 1909206.}}
	\author[1]{Sanjay Goyal}
	\author[1]{Mihaela Beluri}
	\author[2]{Theodore S. Rappaport\vspace{-0.2cm}}
	\affil[1]{InterDigital Communications, Inc., NY, USA, \{sanjay.goyal, mihaela.beluri\}@interdigital.com}
	\affil[2]{NYU WIRELESS, NYU Tandon School of Engineering, Brooklyn, NY, USA, \{ojask, tsr\}@nyu.edu \vspace{-0.4cm}}
	\maketitle 
	\thispagestyle{firststyle}
	\begin{abstract}
Joint communication and sensing allows the utilization of common spectral resources for communication and localization, reducing the cost of deployment. By using fifth generation (5G) New Radio (NR) (i.e., the 3rd Generation Partnership Project Radio Access Network for 5G) reference signals, conventionally used for communication, this paper shows sub-meter precision localization is possible at millimeter wave frequencies. We derive the geometric dilution of precision of a bistatic radar configuration, a theoretical metric that characterizes how the target location estimation error varies as a function of the bistatic geometry and measurement errors. We develop a 5G NR compliant software test bench to characterize the measurement errors when estimating the time difference of arrival and angle of arrival with 5G NR waveforms. The test bench is further utilized to demonstrate the accuracy of target localization and velocity estimation in several indoor and outdoor bistatic and multistatic configurations and to show that on average, the bistatic configuration can achieve a location accuracy of 10.0 cm over a bistatic range of 25 m, which can be further improved by deploying a multistatic radar configuration.
	\end{abstract}
	
	
	\IEEEpeerreviewmaketitle

	\section{Introduction } 
	\label{sec:intro}
As communication systems move towards higher frequency bands, significantly wider system bandwidths become available. The frequency range 2 (FR2) defined in the \gls{3gpp} \gls{5g} \gls{nr} encompasses \gls{mmWave} frequencies from 24.25 to 52.6 GHz and allows a maximum system bandwidth of 400 MHz. As the wireless industry moves towards frequencies above 90 GHz (and eventually Terahertz frequencies) in the future, spectrum spanning several Gigahertz will become available \cite{Rappaport19a,Kanhere2021}. While mobile communication systems evolved with emergency 911 (E-911) capabilities bolted on to the infrastructure to provide position location to within 100 m as part of early second and third generation cellular systems \cite{Rap96a}, future wireless systems will exploit massive bandwidths to provide simultaneous communications and sensing capabilities. Centimeter level localization \cite{Kanhere20a} is one such application that can be made possible by \gls{jcs} without the need of additional sensing hardware, thus reducing cost, control overhead, and power consumption that used to be required to build a separate localization system~\cite{hanzo_JCS_2020, rahman2020enabling,Rap96a}.

For \gls{ue} positioning, the current \gls{3gpp} standards uses measurement techniques, such as \gls{otdoa} and \gls{utdoa}, where the \gls{ue} to be localized must transmit or receive a reference signal \cite{Kanhere20a}. In OTDOA positioning,  the \gls{ue} receives  known reference signals (the \gls{prs}) from multiple \glspl{bs} and the propagation time differences of the reference signal are measured. In UTDOA positioning, the \gls{ue} transmits reference signals (the \gls{srs}) and the propagation time differences are measured at multiple \glspl{bs} \cite{Kanhere20a}. In addition to \gls{ue} positioning, \gls{otdoa} and \gls{utdoa} measurements may enable \gls{jcs} applications such as device-free target localization, wherein the target to be localized is not required to transmit or receive signals. In assisted living facilities, device-free localization could be used for fall detection \cite{Wang_2017}. In \gls{v2x} applications, device-free localization is required for pedestrian avoidance. Security applications such as intrusion detection in homes and offices are made possible via device-free localization without the knowledge or cooperation of the intruder \cite{Youssef_2007}.

Radars have long been used for target detection and localization. With \gls{jcs}, radar functionality can be incorporated into \gls{bs} and \gls{ue}~\cite{rahman2020enabling}. In monostatic radar configurations, a single node (e.g., a \gls{ue} or a \gls{bs} in the network) is used for both transmission and reception. The architecture of a monostatic radar may be half-duplex, wherein the node must transmit and receive over two non-overlapping time intervals, or full-duplex, wherein the node may simultaneously transmit and receive.
Half-duplex monostatic radars may suffer from the drawback that the radar may not be able to sense nearby targets. The minimum radar sensing distance, which is the minimum distance the target (to be localized) must be from the radar transmitter, may be very large. Half-duplex monostatic radars cannot detect target reflections until the transmitted pulse completely leaves the \gls{tx} antenna and the radar switches from \gls{tx} to \gls{rx} mode, during which targets closer than the minimum radar sensing distance cannot be detected. For example, for a 5G NR system at \gls{mmWave} with 120 kHz subcarrier spacing, assuming that the minimum pulse duration is at least  an OFDM symbol duration, i.e., 8.33 us, the minimum radar sensing distance would be 1.25 km due to the speed of radiowave propagation.  

Full-duplex monostatic radars can eliminate the long minimum ranging distance problem; however, the challenge of self-interference must then be addressed. Self-interference may be canceled through active RF cancellation methods, while preserving the reflected target echo \cite{Barneto_2019}. The suitability of LTE reference signals for target localization was analyzed in \cite{Evers_2014}, by using the ambiguity function, which is the output of the matched filter at the \gls{rx} when the known radar waveform is used as the filter input. The LTE reference signals were found to be suitable for full duplex monostatic radar applications, with a range resolution of 7.56 m using a bandwidth of 20 MHz \cite{Evers_2014}. 

Bistatic and multistatic radar configurations, the focus of the rest of the paper, require two or more nodes to operate, wherein full duplex communication is not required because the TX and RX are not collocated. Therefore, bistatic and multistatic configurations are easier to implement with current communication hardware than monostatic configuration. Besides nodes of mobile wireless communication systems (e.g., BS, \gls{ue}) implicitly form bistatic or multistatic configurations with potential targets, hence network nodes may be used for target localization as well as for communications \cite{rahman2020enabling}.

\begin{figure}
		\centering
		\includegraphics[width=0.2\textwidth]{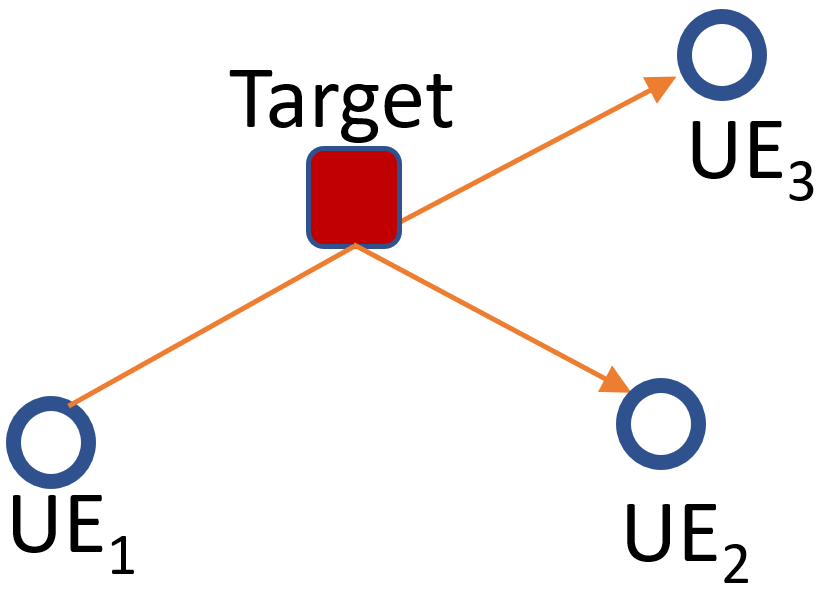}
			\captionsetup{font=small}
		\caption{The co-linear node pair (UE$_1$-UE$_3$) is more sensitive to measurement errors than (UE$_1$-UE$_2$), due to a greater \gls{gdop}. }
		\label{fig:toy_example}
				\vspace{-0.55cm}
	\end{figure}
	
In this work, we analyze the localization accuracy of bistatic and multistatic radar using the 5G NR \gls{dmrs}. We provide an overview of bistatic and multistatic target localization techniques, where an analytical framework is also presented to calculate the \gls{gdop}, a metric to characterize the target location estimation error as a function of the bistatic geometry and the measurement errors. Prior works in \cite{Kong_2016,Lv_2010} calculated the \gls{gdop} with \gls{aoa} and \gls{toa} measurements with bistatic and multistatic radar. In this work we extend such analysis to calculate the \gls{gdop} with \gls{aoa} and \gls{tdoa} measurements with a bistatic radar. We further develop a 5G NR compliant software test bench, consisting of an easy to use software simulation platform in MATLAB, to derive the \gls{tdoa} and \gls{aoa} measurements errors using 5G NR \gls{dmrs} for measurements under different bistatic scenarios, i.e. varying the TX-target-RX separation and target \gls{rcs} in ideal channel conditions. The \gls{aoa} and \gls{tdoa} measurement errors are then used to derive the target location estimation errors for different bistatic as well as multistatic scenarios. We show that for a given target, the \gls{gdop} can be used as a metric by which to select the best nodes for localization, as well as the mode of operation of each node (transmit or receive). For instance, as seen in Fig. \ref{fig:toy_example}, for the given target, bistatic measurements using nodes UE$_1$ and UE$_2$ may be preferred over the measurements using nodes UE$_1$ and UE$_3$ for target localization due to lower \gls{gdop}. We explicitly derive the \gls{gdop} in Section \ref{sec:gdop}.

The rest of the paper is organized as follows: Section \ref{sec:sec2} provides an overview of bistatic and multistatic target localization techniques along with the details of \gls{gdop}. The details of 5G NR compliant test bench used to derive the \gls{aoa} and \gls{tdoa} measurement errors are given in Section~\ref{sec:sec3}. Section \ref{sec:sec4} provides simulation results of target location estimation errors for various bistatic and multistatic scenarios. Section~\ref{sec:conc} concludes the paper.
\begin{figure}
		\centering
		\includegraphics[width=0.45\textwidth]{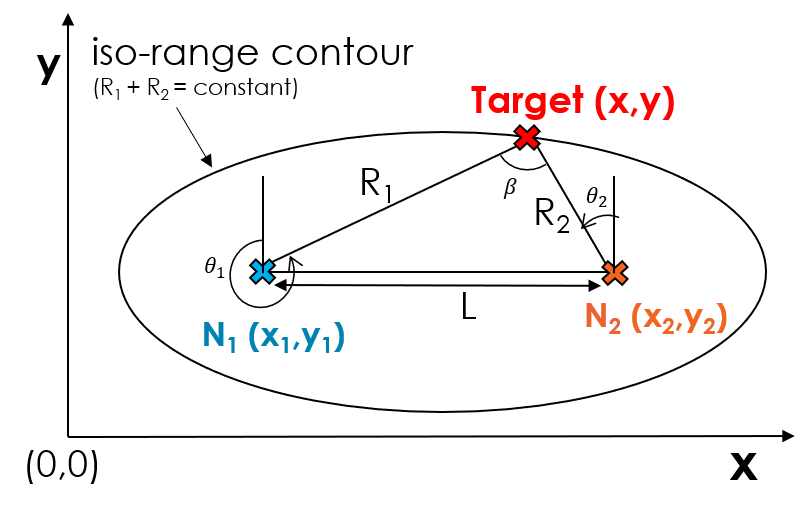}
			\captionsetup{font=small}
		\caption{Illustration of the bistatic radar geometry, where  the position of the target is determined using the \gls{aoa} and \gls{tdoa} measurements.}
		\label{fig:bistatic_geometry}
		\vspace{-0.55cm}
	\end{figure}
	
	\section{Localization in Bi/Multi-static Radar Configuration }
	\label{sec:sec2}
	\subsection{Bistatic Localization Geometry}\label{sec:bistatic_geometry}
	The bistatic radar configurations employs two nodes, $ N_1 $ and $ N_2 $, at different known locations $(x_1,y_1)$, and $(x_2,y_2)$, respectively. A known signal, called a reference signal, may be transmitted from either node. The unknown position of the target is determined by measuring the \gls{tdoa} and \gls{aoa} of the reference signal at the \gls{rx} node. The bistatic radar has two modes of operation --- in mode 1, $ N_1 $ transmits the reference signal to $ N_2 $ and in mode 2, $ N_2 $ transmits the reference signal to $ N_1 $. We now discuss how the position of the target is determined in mode 1, and it will be clear that a symmetric procedure is applicable to mode 2. 
	
	In mode 1, the bistatic \gls{rx} ($ N_2 $) measures the time delay between the direct signal from the \gls{tx} ($ N_1 $) and the signal reflected off the target, i.e., the \gls{tdoa} $\Delta T$:
	\begin{align}
	\Delta T = \dfrac{R_1+R_2}{c}-\dfrac{L}{c},
	\end{align}
	where $R_1$ is the distance of the target from $ N_1 $, $R_2$ is the distance of the target from $ N_2 $, L is the distance between $ N_1 $ and $ N_2 $, and $c$ is the speed of light. Target locations where $R_1+R_2$ is constant lie on an ellipse (with focii at $N_1 $ and $ N_2 $) called the iso-range contour, as shown in Fig \ref{fig:bistatic_geometry}. $ N_2 $ also measures the \gls{aoa} ($\theta_2$) of the reflected signal. Using the \gls{tdoa} and \gls{aoa}, $R_2$ can be calculated as follows \cite{Johnsen_2005}:
	\begin{align}
	R_2 = \dfrac{c^2\Delta T^2 + 2 c L \Delta T}{2\left(\left(c \Delta T + L)\right) - L \sin{\theta_2}\right)}.
	\end{align}
	
	Finally, the position of the target ($x,~y$) is estimated to be $(x_2 - R_2 \sin \theta_2,~y_2 + R_2 \cos \theta_2$). 
	
	The \gls{tdoa} and \gls{aoa} estimation performance of a bistatic radar depends on the \gls{snr} at the \gls{rx}, given by \cite{Kuschel_2019}:
	\begin{align}
	SNR = \dfrac{P_T G_T G_R \lambda^2 \sigma_B}{\left(4 \pi\right)^3 \left(R_1 R_2\right)^2 k T_s B},
	\end{align}
	where $P_T$ is the \gls{tx} power in watts, $G_T$ and $G_R$ are the \gls{tx} and \gls{rx} antenna gains, $\lambda$ is the wavelength of the transmitted signal, $\sigma_B$ is the bistatic \gls{rcs} of the target, k is the Boltzmann constant ($1.38\times 10^{-23}$ J/K), $T_s$ is the \gls{rx} noise temperature, and $B$ is the noise bandwidth at the \gls{rx}. Note that constant SNR curves (called \textit{Cassini curves}) do not coincide with iso-range contour, i.e.,  the \gls{snr} received at the \gls{rx} varies for different target locations on an iso-range~\cite{Kuschel_2019}.
	
	\subsection{Bistatic target localization \gls{gdop} with \gls{tdoa} and \gls{aoa} measurements}\label{sec:gdop}
	The \gls{gdop} expresses the sensitivity of the target location estimate to measurement errors, (e.g., errors in \gls{tdoa}, \gls{aoa}) as well as errors in the known location of $N_1$ and $N_2$. High \gls{gdop} indicates that the radar geometry is very sensitive to error in measurements – a slight error in measurements leads to large position location error. A low \gls{gdop} is preferred, where the position location error is insensitive to measurement errors. 

	GDOP has been calculated for various radar localization methods, such as systems where only \gls{toa} is measured or for hybrid \gls{toa}/\gls{aoa} localization methods \cite{Kong_2016}. Although the \gls{gdop} of joint \gls{aoa} and \gls{tdoa} has been calculated using two transmitting reference nodes \cite{Liu_2013}, to the best of our knowledge, the \gls{gdop} of a bistatic radar measuring  \gls{aoa} and \gls{tdoa} for device-free passive localization has not been examined, as shall now be derived. 
	
The measured \gls{tdoa} and \gls{aoa} can be expressed in terms of the coordinates of the target, $ N_1 $, and $ N_2 $ as follows \cite{Kong_2016}:
	\begin{align}
	\nonumber
&\Delta T =	f_{\Delta_T}(x,y) = \dfrac{1}{c}[R_1 + R_2-L]\\\nonumber
	&=\dfrac{1}{c} \bigg[\sqrt{(x-x_1)^2+(y-y_1)^2}\nonumber
	+\sqrt{(x-x_2)^2+(y-y_2)^2} \\
	&- \sqrt{(x_1-x_2)^2+(y_1-y_2)^2}\bigg], \label{eq:deltaT}\\
	\theta_i&=f_{\theta_i}(x,y)	 = \tan^{-1}\left(\dfrac{x_i-x}{y-y_i}\right),\label{eq:theta_2}
	\end{align}
	where $i = 1,2$ for the \gls{aoa} measurement at node $ N_1 $ and $ N_2 $, respectively. 
	
	Taking the total derivative of the measurements, $\Delta T$, $\theta_1$, and $ \theta_2 $ with respect to the coordinates of the target, $N_1$, and $N_2$ we find:
	\begin{align}
	&\dfrac{\partial \Delta T}{\partial x} = \dfrac{1}{c}\left[ \dfrac{x-x_1}{R_1}+ \dfrac{x-x_2}{R_2}\right], \\
	&\dfrac{\partial \Delta T}{\partial y} = \dfrac{1}{c}\left[ \dfrac{y-y_1}{R_1}+ \dfrac{y-y_2}{R_2}\right],\\
	&\dfrac{\partial \Delta T}{\partial x_1} = \dfrac{1}{c}\left[ \dfrac{x_1-x}{R_1}- \dfrac{x_1-x_2}{L}\right],
			\end{align}
		\begin{align}
	&\dfrac{\partial \Delta T}{\partial y_1} = \dfrac{1}{c}\left[ \dfrac{y_1-y}{R_1}- \dfrac{y_1-y_2}{L}\right], 	\\
	&\dfrac{\partial \Delta T}{\partial x_2} = \dfrac{1}{c}\left[ \dfrac{x_2-x}{R_2}- \dfrac{x_2-x_1}{L}\right],\\
	&\dfrac{\partial \Delta T}{\partial y_2} = \dfrac{1}{c}\left[ \dfrac{y_2-y}{R_2}- \dfrac{y_2-y_1}{L}\right].
	\end{align}
	
	Let  $ v_i = \dfrac{x_i-x}{y-y_i}$, where $i = 1,2$
	\begin{align}
	&\dfrac{\partial \theta_i}{\partial x} = -\dfrac{1}{1+v_i^2}\left[ \dfrac{1}{y-y_i}\right] =-\dfrac{\partial \theta_i}{\partial x_i}, \\
	&\dfrac{\partial \theta_i}{\partial y} = -\dfrac{v_i}{1+v_i^2}\left[ \dfrac{1}{y-y_i}\right] =-\dfrac{\partial \theta_i}{\partial y_i}, \\
	&\dfrac{\partial \theta_1}{\partial x_2}=\dfrac{\partial \theta_1}{\partial y_2}=\dfrac{\partial \theta_2}{\partial x_1}=\dfrac{\partial \theta_2}{\partial y_1}=0.
	\end{align}
	
	For small errors, the differentials of the target position $(dx,dy)$, \gls{tdoa} ($d\Delta T$), \gls{aoa} ($d\theta_1$ and $ d\theta_2 $), node positions $(dx_1,dy_1)$ and $(dx_2,dy_2)$ are approximately equal to the estimation errors \cite{Kong_2016}.
	Let $\mathbf{dp} = [dx$ $dy]^T$, and $\mathbf{dX}  = [dx_1$ $dy_1$ $dx_2$ $dy_2]^T$ denote the estimation errors of target position, and node position in vector form, respectively. We now evaluate the relation between $\mathbf{dp}$ and the geometry of the nodes.
	
	In mode 1, since $ \theta_2 $ is measured at $N_2$, let
	$\mathbf{Z}  = [\Delta T~\theta_2 ]^T$ denote the measurements in vector form. Taking the total derivative of the measurement vector, we obtain:
	\begin{align}
	\mathbf{dZ}  = \mathbf{C_1}  \mathbf{dp} + \mathbf{C_2}  \mathbf{dX} ,
	\end{align}
	where
	\begin{align}
	\mathbf{C_1}  &= \begin{bmatrix} 
	\dfrac{\partial \Delta T}{\partial x} & \dfrac{\partial \Delta T}{\partial y}  \\
	\dfrac{\partial \theta_2}{\partial x} & \dfrac{\partial \theta_2}{\partial y} 
	\end{bmatrix}, \\
	\mathbf{C_2}  &= \begin{bmatrix} 
	\dfrac{\partial \Delta T}{\partial x_1} & \dfrac{\partial \Delta T}{\partial y_1} &\dfrac{\partial \Delta T}{\partial x_2} & \dfrac{\partial \Delta T}{\partial y_2} \\
	\dfrac{\partial \theta_2}{\partial x_1} & \dfrac{\partial \theta_2}{\partial y_1} &\dfrac{\partial \theta_2}{\partial x_2} & \dfrac{\partial \theta_2}{\partial y_2} 
	\end{bmatrix}. 	
	\end{align}
	Thus,
	\begin{align}
	\label{eq:dp}
	\mathbf{dp} &= (\mathbf{C_1} ^T\mathbf{C_1} )^{-1}\mathbf{C_1} ^T(\mathbf{dZ} -\mathbf{C_2}  \mathbf{dX} ),\\ 
	\nonumber
	\mathbf{P_{dp}} &= E[\mathbf{dp} \mathbf{dp}^T] 	\\
	&= \mathbf{B}  \{E[\mathbf{dZ}  \mathbf{dZ} ^T]+\mathbf{C_2}  E[\mathbf{dX}  \mathbf{dX} ^T]\mathbf{C_2} ^T\}\mathbf{B} ^T,
	\end{align}
	where $\mathbf{P_{dp}}$ is the error covariance matrix, $ \mathbf{B} =  (\mathbf{C_1} ^T\mathbf{C_1} )^{-1}\mathbf{C_1} ^T $, $ E[\mathbf{dZ}  \mathbf{dZ} ^T] $ = diag($ \sigma_{\Delta T}^2 $, $ \sigma_{\theta_2}^2 $) assuming the \gls{tdoa} and \gls{aoa} measurement errors are uncorrelated, and $ E[\mathbf{dX}  \mathbf{dX} ^T] $ = diag($\sigma_{x_1}^2 $, $ \sigma_{y_1}^2 $, $ \sigma_{x_2}^2 $, $ \sigma_{y_2}^2 $) assuming the errors in the node positions are uncorrelated. $ \sigma_{\Delta T}^2 $ and $ \sigma_{\theta_2}^2 $ are the variances in measured \gls{tdoa} and \gls{aoa} respectively, while $\sigma_{x_1}^2 $, $ \sigma_{y_1}^2 $, $ \sigma_{x_2}^2 $, and $ \sigma_{y_2}^2 $ are the variances in the coordinates of $N_1$ and $N_2$. Note that $\mathbf{dp}$, the target position estimation error, is a function of the position of the target on the iso-range contour with different points on the contour having different target location error even though the \gls{tdoa} and \gls{aoa} errors are the same. Finally, 
	\begin{align}
	GDOP = \sqrt{trace(\mathbf{P_{dp}} )}. 
	\end{align}
	
	The GDOP can be similarly derived for mode 2 by by replacing $ \theta_2 $ with $ \theta_1$ in the definitions of $ \mathbf{Z}  $, $ \mathbf{C_1}  $ and $ \mathbf{C_2}  $. 

\subsection{Extension to Multistatic Configurations}	\label{sec:multistatic}
In scenarios when there are more than two nodes capable of performing \gls{tdoa} and \gls{aoa} measurements from other reference nodes, the position of the target may be obtained by estimating the target position for different node pairs and then combining the estimates in an intelligent manner. Measurements at the nodes may be treated as soft information \cite{Conti_2019}. The measurements from each node pair may be sent to a central server (or to one of the nodes involved in the measurements), where a single position estimate is obtained using soft information from all the nodes. One method to determine the target position $ (x,y)$ is to minimize $L(x,y)$, the weighted least squares loss metric defined as \cite{Al_Jazzar_2009}:
\begin{align}
\nonumber
L(x,y) =	\sum_{i=1}^{N}w_i\Big[\Big(a_i(c \cdot (\Delta T_{i,meas}-f_{\Delta T_i}(x,y))\Big)^2\label{eq:nls}\\
	+\Big(b_i(\theta_{2_{i,meas}}-f_{\theta_{2_i}}(x,y))/2\pi)\Big)^2 \Big]
\end{align}
where there are $ N $ node pairs conducting measurements, $a_i$ and $b_i$ are weighting parameters assigned to TDOA and AOA measurements, respectively ($a_i$ has units $m^{-1}$ while $b_i$ is unit-less),  $\Delta T_{i,meas} $ and $ \theta_{2_{i,meas}} $ are the measured \gls{tdoa} and \gls{aoa} at the $ i^{th} $ node pair, respectively, while  $ f_{\Delta T_i}(x,y) $ and $ f_{\theta_2}(x,y) $ are defined as in (\ref{eq:deltaT}) and (\ref{eq:theta_2}), respectively, and $w_i$ are the weights assigned to each node pair, inversely proportionate to the predicted positioning error of the node pair ($|\mathbf{dp_i}|$). Weighted averaging with $w_i$ ensures that node pairs with
poor geometry (high $\mathbf{dp}$) will have a lower impact
on the position estimate. The optimization problem may be solved by techniques such as the Levenberg–Marquardt algorithm \cite{More_1978}.

	\section{Bi/Multi-static Configuration with 5G NR }\label{sec:sec3}
\begin{figure}
		\centering
		\includegraphics[width=0.48
		\textwidth]{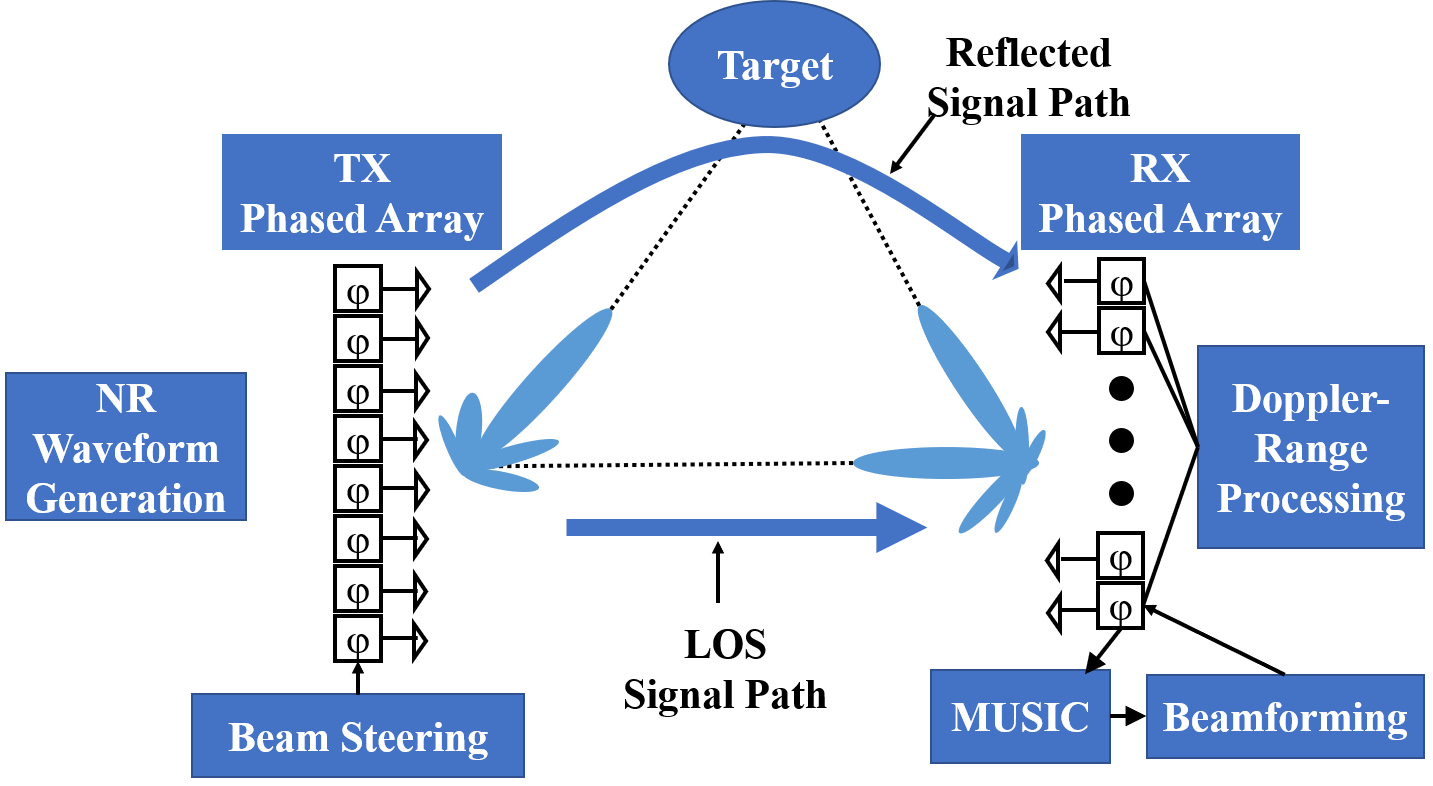}
			\captionsetup{font=small}
		\caption{The 5G NR test bench set-up depicting how the generated waveform is steered by the \gls{tx} phased array towards the target. MUSIC at the RX estimates the \gls{aoa} of the signal reflected from the target, which controls the \gls{rx} beamforming direction.}
		\label{fig:architecture}
		\vspace{-0.55cm}
	\end{figure}

	Before real-world deployment, link-level simulations help determine the feasibility of localization solutions. With the help of the Phased Array MATLAB toolbox, we developed a 5G NR compliant test bench to quantify the measurements errors in \gls{tdoa} and \gls{aoa} for different bistatic radar configurations in order to evaluate the localization performance of a 5G NR system with radar capabilities, as illustrated in Fig. \ref{fig:architecture}.
	
	Highly directional antenna arrays used at \gls{mmWave} frequencies have narrower beamwidth; to ensure that the direct signal and the reflected signal are received at the \gls{rx} with adequate strength, the \gls{rx} needed to use two RF chains to simultaneously beamform in two directions, one towards the target, and the other towards the \gls{tx}.  An \gls{rx} with one RF chain which can steer towards only one direction at a time, may first steer towards the \gls{tx} to receive the direct signal, and then after a fixed time interval (an integral multiple of the periodicity of the reference signal transmission) steer towards  the target to receive the reflected signal. The fixed time interval is subtracted from the time at which the reflected signal arrives at the \gls{rx} to derive the TDOA.

	In real-world bistatic radar implementations, the \gls{tx} antenna array is electrically swept across different steering angles since the direction of the target is not known. However, since this work focuses on quantifying the performance of the radar system and not the performance of the beamsweeping algorithm, we assumed that the \gls{tx} was steered directly towards the target. The \glspl{aoa} of the signals at the \gls{rx} were estimated via the MUSIC algorithm \cite{Schmidt_1986,Rappaport_1999}.
	
	The \gls{tdoa} of the direct signal and the reflected signal was estimated via matched filtering, wherein the received waveform was correlated with a time-reversed copy of the transmitted waveform. The \gls{tdoa} estimate is the time delay between the peaks in the correlation function. To ensure that the reflected signal power of weak targets received using the \gls{rx} beam directed towards the target was not swamped by the direct signal, a scaled copy of the signal received from the direct path using the \gls{rx} beam directed towards the \gls{tx}  was removed from the reflected signal received by the \gls{rx}.
	
	To measure the Doppler shift in the received signal, the NR waveform generated by the \gls{tx} was repeatedly transmitted as a train of 64 waveforms, which were each passed through a matched filter at the \gls{rx}. The \gls{rx} generated a Range-Doppler plot by calculating the discrete Fourier transform (DFT) of the matched filter response of each waveform and determining the common Doppler frequency present in all waveforms \cite{Koks_2014}. If the Doppler frequency measured by the bistatic radar is $f_D$, the velocity of the target is given by \cite{Kuschel_2019}:
	\begin{align}
	v = \dfrac{d}{dt}(R_1+R_2) = c \dfrac{f_D}{f_0},
	\end{align}
	where c is the speed of light and $f_0$ is the carrier frequency. Alternatively, a Kalman filter could be used to track a moving target and estimate the velocity\cite{Kanhere_2021_b}.

	When radar clutter \cite{rahman2020enabling}, multipath, and interference from other \glspl{tx} (currently not modeled in the test bench) are considered, the localization performance will likely be degraded.

	\section{Performance Evaluation}\label{sec:sec4}
	
		As described in Section~\ref{sec:bistatic_geometry}, the target localization algorithms used in this paper rely on two types of measurements --- \gls{aoa} and \gls{tdoa}. Therefore, we first simulated the error of both measurements using the 5G NR test bench and then used the mean absolute error of the simulated measurements to evaluate the target localization accuracy. 
		
	The target position and velocity estimation simulations were conducted at 28 GHz using the \gls{dmrs} as the reference signal. To emulate target localization in a real NR system, randomly generated data was transmitted along with \gls{dmrs} (i.e., on resource elements not occupied by \gls{dmrs}), however it was assumed that the \gls{rx} had no knowledge of the transmitted data and that the matched filter at the \gls{rx} was designed to match the \gls{dmrs} signal alone. A single \gls{dmrs} symbol was transmitted per time slot (\gls{dmrs} \textit{Additional Position} = 0), with six subcarriers utilized per resource block (\gls{dmrs} \textit{Configuration Type-I})~\cite{TS38211}. Since no impact was observed by simulating other \gls{dmrs} configurations (i.e., with different density of \gls{dmrs} in time or/and frequency), we only present the results with \gls{dmrs} \textit{Configuration Type-I}. 
	
	A broader antenna beamwidth at the \gls{tx} compared to the \gls{rx} ensured that the \gls{los} signal path did not fall in a deep antenna null when the \gls{tx} was steered towards the target. Hence, the \gls{tx} antenna had 8 while the \gls{rx} antenna had 16 elements. The maximum  effective  isotropic radiated  power was set to 43 dBm and the RX noise figure was set to 13 dB~\cite{3GPP.38.101.2}. A SCS of 120 kHz was chosen for the simulations with a bandwidth of 100 MHz and 400 MHz.

	An analysis of the \gls{tdoa} and \gls{aoa} errors was conducted for three bistatic scenarios with different bistatic ranges ($R_1+R_2-L$) to simulate different short and mid range bistatic applications, for example, localization for VR gaming, first responders indoors, and vehicles outdoors. Scenario 1 with L = 3 m and $ R_1+R_2 = 6 $ m, scenario 2 with L = 15 m and $ R_1+R_2 = 30 $ m, and scenario 3 with  L = 25 m and $ R_1+R_2 = 50 $ m were simulated. Increasing $L$ further resulted in very low SNR at the \gls{rx} node, in which case MUSIC for \gls{aoa} estimation would fail.
	
	The target was assumed to be isotropic, i.e., a constant bistatic \gls{rcs} for all orientations. The bistatic \gls{rcs} of the target was set to -20 dBm$ ^2 $ in scenario 1 in order to model a target with low scattering power (e.g., a person)  \cite{Motomura_2018}, and 0 dBm$ ^2 $ in scenario 2 and 3 to model a more reflective target (e.g., vehicles) \cite{Motomura_2018}. For localization accuracy evaluation, the target, TX, and RX were assumed to be stationary.

	\subsection{\gls{tdoa} and \gls{aoa} Estimation Errors}
\begin{figure}
		\centering
		\includegraphics[width=0.5\textwidth]{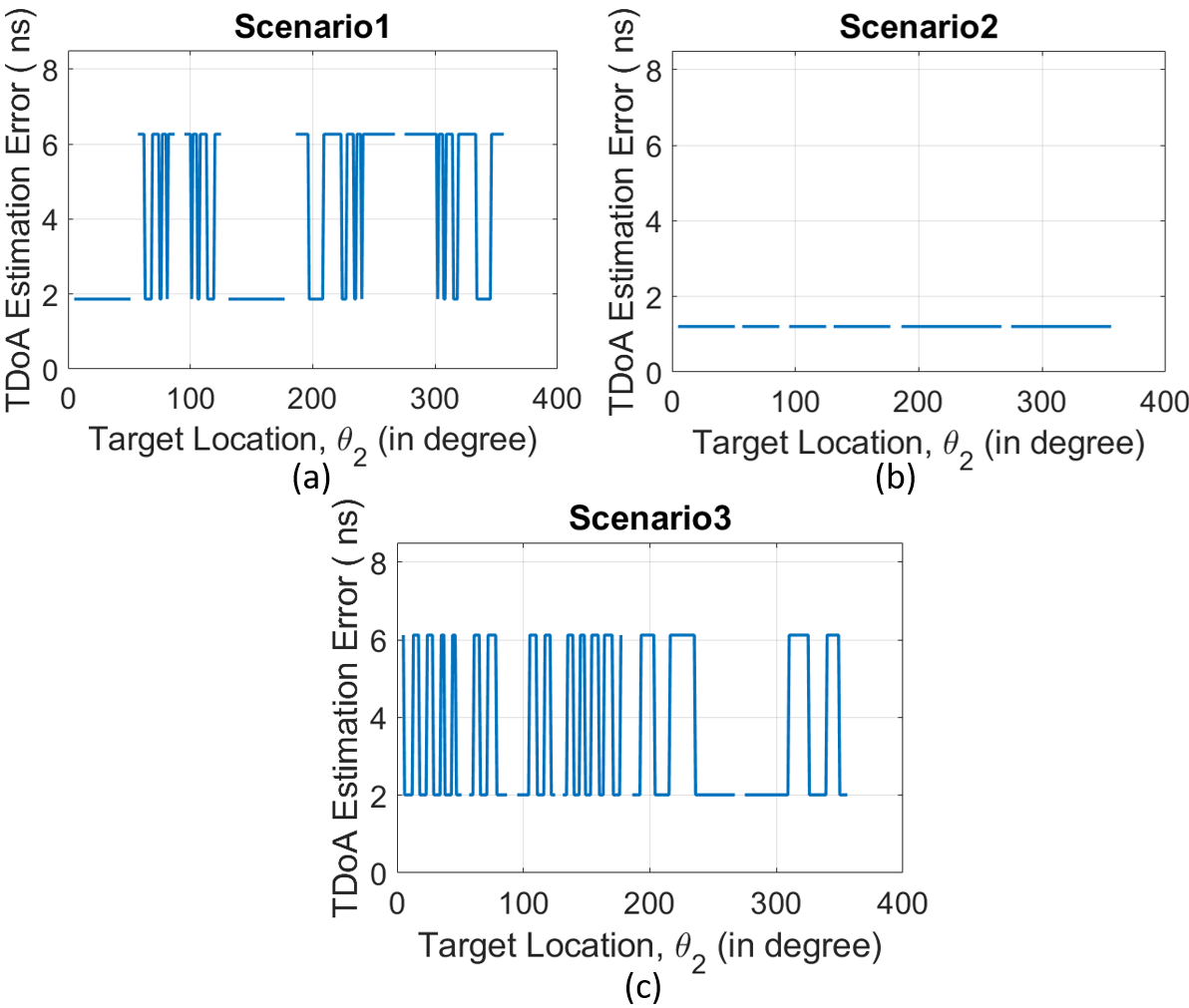}
			\captionsetup{font=small}
		\caption{The TDOA estimation error for target localization along the iso-range contour with signal BW = 100 MHz.}
		\label{fig:TDoA}
		\vspace{-0.2cm}
		\end{figure}

\gls{tdoa} estimation errors for the three scenarios are plotted in Fig. \ref{fig:TDoA} for different points on the iso-range contour with RF bandwidth of 100 MHz. The main source of error for \gls{tdoa} measurements was the finite resolution of the measured time. The true \gls{tdoa} in scenario~1 was 10.0 ns. Due to the finite \gls{tdoa} resolution of 8.14 ns (assuming an I/Q sampling rate of 122.88 MHz when the signal bandwidth is 100 MHz), at different points on the iso-range contour, the \gls{tdoa} was estimated to be 8.14 ns or ($8.14 \times 2$ =) 16.28 ns (integral multiples of the \gls{tdoa} resolution), due to which \gls{tdoa} error oscillated between 1.86 ns and 6.28 ns in Fig. \ref{fig:TDoA}(a).
 
 Similarly in scenario 2, the true \gls{tdoa} is 50 ns, yet the radar estimates the \gls{tdoa} to be one quantization step below the true value, at ($8.14\times 6$ =) 48.84 ns, due to which an error of 1.16 ns is observed in Fig. \ref{fig:TDoA}(b). Finally, in scenario 3, the true \gls{tdoa} is 83.4 ns which falls between ($8.14\times 10$ =) 81.4 ns and ($8.14\times 11$ =) 89.5 ns, due to which the \gls{tdoa} error oscillated between 2.0 ns and 6.1 ns in Fig. \ref{fig:TDoA}(c).
 
 Increasing the RF bandwidth from 100 MHz to 400 MHz improved the \gls{tdoa} estimation accuracy as the minimum time resolution improved from 8.14 ns to 2.03 ns. The absolute mean \gls{tdoa} errors (over the corresponding iso-range contour) were 4.2 ns, 1.21 ns, and 3.55 ns for scenarios 1, 2, and 3, respectively, when a RF bandwidth of 100 MHz was simulated. Increasing the bandwidth to 400 MHz, reduced the  absolute mean \gls{tdoa} errors to 0.17 ns, 1.21 ns, and 0.02 ns. 
	
		\begin{figure}
	\centering
	\includegraphics[width=0.5\textwidth]{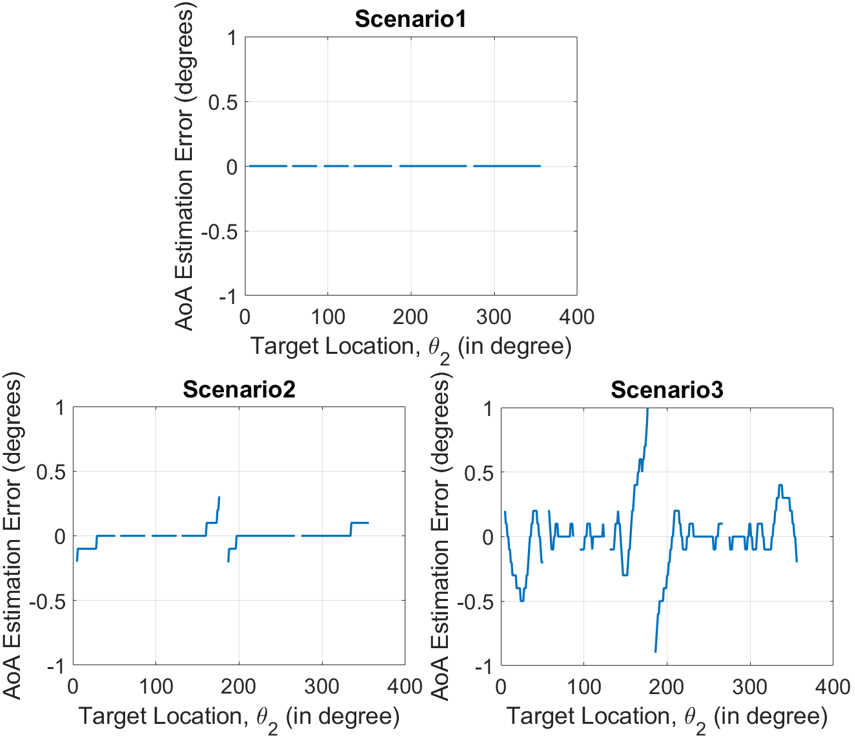}
		\captionsetup{font=small}
	\caption{The AOA estimation error for target localization along the iso-range contour with signal BW = 100 MHz.}
	\label{fig:AoA}
	\vspace{-0.6cm}
\end{figure}
	
MUSIC was able to very accurately estimate the \gls{aoa}, as seen in Fig. \ref{fig:AoA} (with bandwidth of 100 MHz), with sub-degree accuracy achieved at all points on the iso-range contour. As the distance between the target and the nodes increased, the performance of MUSIC deteriorated due to the worsening SNR, however sub-degree accuracy was achieved at all points on the iso-range curve for all three scenarios, with nearly no \gls{aoa} error for scenario 1 with nodes 3 m apart.

The discontinuities in the error plots of Fig. \ref{fig:TDoA} and \ref{fig:AoA} occur since the \gls{tdoa} and \gls{aoa} could not be measured when the target, the \gls{tx} and the \gls{rx} were collinear (at $ \theta_2 $ = 90$ ^\circ $, 270$^\circ  $) as the \gls{los} signal would completely swamp the signal reflected from the target. The boresight of the phased arrays was oriented along the line joining the two nodes. The \gls{tdoa} and \gls{aoa} of targets perpendicular to boresight could not be determined due to antenna nulls directed towards the target.

As RF bandwidth was increased from 100 MHz to 400 MHz, a slight decrease in \gls{aoa} accuracy was observed since the noise power ($kTB$) increased by 6 dB. For example, the absolute mean \gls{aoa} (over the corresponding iso-range contour) were 0$^\circ$, 0.03$^\circ $, and 0.16$ ^\circ $ for scenarios 1, 2, and 3, respectively with bandwidth of 100 MHz, which became 0$^\circ$, 0.03$^\circ $, and 0.23$ ^\circ $, respectively with bandwidth of 400 MHz.
		
\subsection{Utilizing TDOA and AOA estimation errors to evaluate localization performance}
To evaluate the target localization performance, the mean absolute \gls{aoa} and \gls{tdoa} errors over all points on the iso-range contour were used. The error in the known locations of $N_1$ and $N_2$ was assumed to be 0.01 m. A low error in the TX/RX locations was selected to focus on the effect of errors in \gls{aoa} and \gls{tdoa} measurements. The location estimation errors after taking the measurement errors into account were evaluated for all three scenarios via the bistatic localization technique described in Section \ref{sec:bistatic_geometry} for both modes of radar operation. By channel reciprocity, the mean absolute measurement errors were equal for both modes of operation.

\begin{figure}
	\centering
	\includegraphics[width=0.40\textwidth]{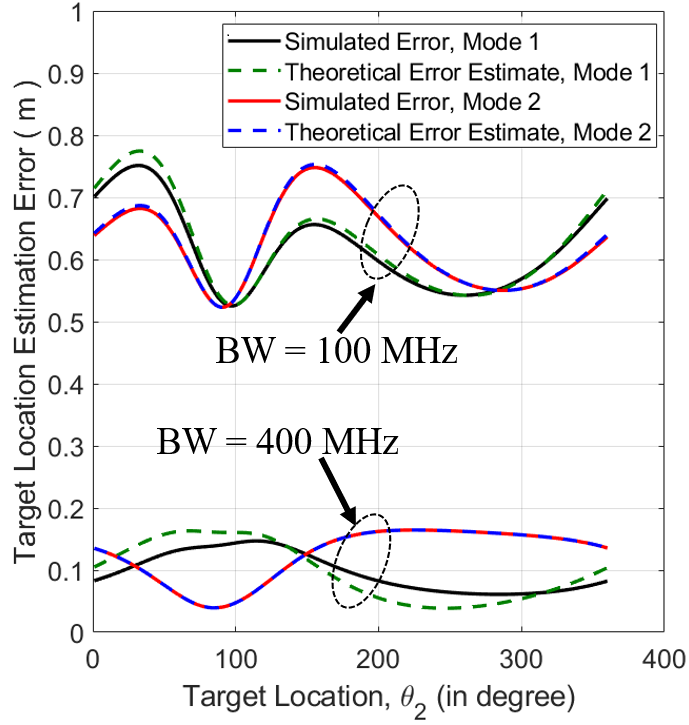}
	\captionsetup{font=small}
	\caption{The simulated position location error for a bistatic configuration is predicted well by (\ref{eq:dp}). Hence, \gls{gdop} may be used for mode selection. The parameters used to simulate the localization error correspond to scenario 3 (with a bistatic range of 25 m), with node location errors of 0.01 m.}
	\label{fig:bistatic_error}
	\vspace{-0.45cm}
\end{figure}

The simulation results for scenario 3 are depicted in Fig. \ref{fig:bistatic_error}, illustrating how the relative distances and angles between the target and the nodes affects the location estimation errors with the same measurement errors, as noted in Section \ref{sec:gdop}. Although increasing the bandwidth from 100 MHz to 400 MHz slightly decreased the \gls{aoa} accuracy,  the increase in time resolution had a greater impact on position location accuracy, which improved at 400 MHz. In mode 1, the mean localization error dropped from 0.62 m to 0.10 m when the bandwidth was increased from 100 MHz to 400 MHz. In mode 2, the mean localization error dropped from 0.65 m to 0.12 m. The results for the other scenarios follow a similar trend and are omitted due to lack of space. The error predicted from GDOP analysis (using (\ref{eq:dp})) is also plotted in Fig. \ref{fig:bistatic_error}. As seen in Fig. \ref{fig:bistatic_error}, there is a clear preference of one mode of radar operation over another at different points on the iso-range contour. Since there is very good agreement between the localization error and the error predicted from the GDOP analysis, for a given target location on the iso-range, the GDOP analysis may be used to determine the direction of the transmission ($N_1$ to $N_2$ or $N_2$ to $N_1$) that minimizes the target localization error. Further, if more than two nodes are available for target localization, the set of nodes with lowest predicted \gls{gdop} could be selected.

\begin{figure}
	\centering
	\includegraphics[width=0.35\textwidth]{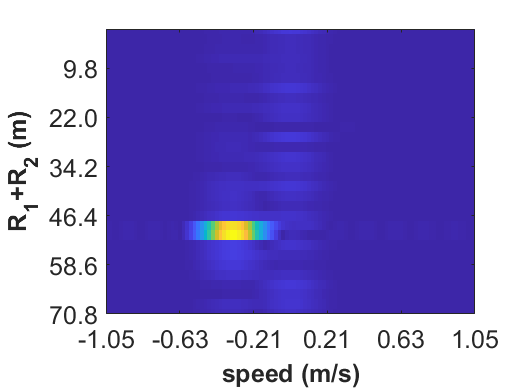}
		\captionsetup{font=small}
	\caption{The position and velocity of the target are simultaneously estimated using the Doppler-range plot of the target in scenario 3.}
	\label{fig:doppler_velocity}
	\vspace{-0.5cm}
\end{figure}

The bistatic radar determined the target velocity with very high accuracy, as illustrated in Fig. \ref{fig:doppler_velocity}. The target, located on the iso-range contour at an angle of $ \theta_2=60^\circ $ was moving in the direction normal to the tangent of the iso-range contour, radially inwards at a speed of 0.2 m/s. The velocity estimation error was 0.008 m/s in scenario 3, with a similar performance observed for the other scenarios.

\begin{figure}
	\centering
	\includegraphics[width=0.35\textwidth]{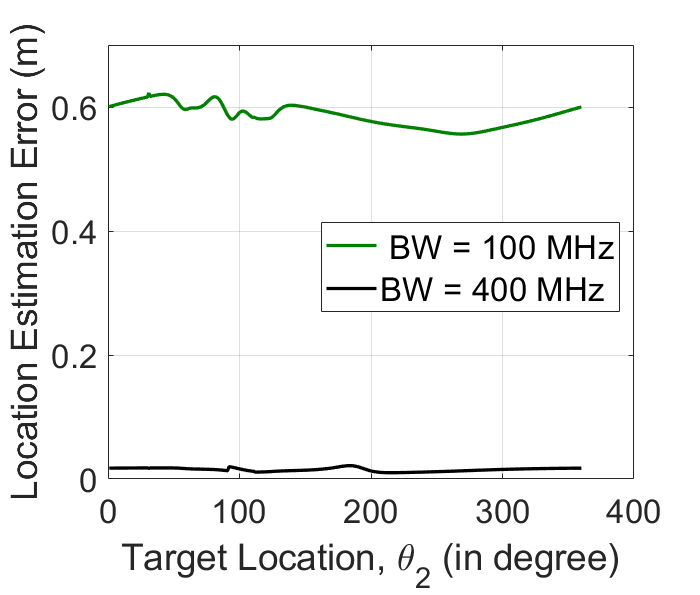}
		\captionsetup{font=small}
	\caption{Multistatic weighted least squares localization performed better when a wider bandwidth of 400 MHz was used.}
	\label{fig:multistatic_errors}
	\vspace{-0.5cm}
\end{figure}

The localization accuracy with the weighted least squares multistatic localization technique described in Section \ref{sec:multistatic} was also evaluated for a multi-static configuration using the mean absolute errors determined from the 5G test bench with system bandwidths of 100 MHz and 400 MHz. One \gls{tx} node and three \gls{rx} nodes were used to localize the target. The \gls{rx} nodes were uniformly placed on a circle with radius $L$  centered at the \gls{tx}. The weighting factors $a_i$ and $b_i $ in (\ref{eq:nls}) were set to 1. A comparison of performance of the weighted least squares algorithms in scenario 3 (with a bistatic range of 25 m) is provided in Fig. \ref{fig:multistatic_errors} (similar trend was observed for other scenarios). The mean localization error was 0.58 m and 0.02 m, respectively, when a signal bandwidth of 100 MHz and 400 MHz was used. 
The results suggest that multistatic localization outperforms bistatic localization, particularly for scenarios with good node geometry (low GDOP) and at wider bandwidths. Hence, the RX must share the measurements (TDOA and AOA) with a central processing unit (e.g., located at the TX), rather than individually calculating a bistatic target estimate. The localization performance was limited by TX-RX location estimation error at 400 MHz (0.01 m), while the performance was limited by the error in the AoA estimates (0.16$^\circ$) and TDoA estimates (3.55 ns) at 100 MHz.
	\section{Conclusions and Future Work} \label{sec:conc}
	With 5G NR, 3GPP has specified operation at mmWave frequencies (FR2). The large available bandwidths in these bands enable both very high data rate communications, and enhanced positioning and localization. This paper focuses on the localization aspect, and provides an analytical method to select the bi/multi-static configuration for a given target location, to minimize the location estimation error. Using simulations with a 5G NR test bench operating at 28 GHz, we evaluated the TDOA and AOA estimation errors for various deployments in ideal channel conditions.  We then used the mean TDOA and AOA estimation errors to estimate the target location in a bistatic configuration, for all target locations on the iso-range, and compared it to the analytical prediction. The good alignment between the simulation and the theoretical analysis indicates that GDOP may be used to select the nodes used in the bistatic configuration for target localization.
	
For example, in a bistatic configuration with $R_1+R_2$~=~50~m and baseline L = 25 m, with TDOA and AOA errors derived from the 5G NR simulation, the mean localization error over the iso-range was 0.62 m (0.10 m) with signal bandwidth of 100 MHz (400 MHz).  We observed that the minimum TDOA resolution (driven by system bandwidth) limits the localization performance.  Lastly, for the same scenario, we showed that the multi-static configuration (using least squares) reduced the localization error to 0.02 m. Future work may include evaluating the performance of other 5G NR reference signals, utilizing other TDOA estimation techniques such as channel estimation based approaches \cite{Wang_2017}, and investigating the effect of radar clutter and multipath on the localization error.

	\bibliography{references}
	\bibliographystyle{IEEEtran}

\end{document}

%% file: acronyms.tex
\newacronym{3gpp}{3GPP}{3rd Generation Partnership Project}
\newacronym{5g}{5G}{fifth generation}
\newacronym{aoa}{AOA}{Angle of Arrival}
\newacronym{aod}{AOD}{Angle of Departure}
\newacronym{bs}{BS}{base station}
\newacronym{eirp}{EIRP}{effective isotropic radiated power}
\newacronym{gnb}{gNB}{next generation NodeB}
\newacronym{dmrs}{DMRS}{demodulation reference signal}
\newacronym{iot}{IoT}{internet of things}
\newacronym{jcs}{JCS}{joint communication and sensing}
\newacronym{los}{LOS}{line-of-sight}
\newacronym{lpp}{LPP}{4G Positioning Protocol}
\newacronym{lte}{LTE}{Long Term Evolution}
\newacronym{ml}{ML}{machine learning}
\newacronym{mmimo}{mMIMO}{massive MIMO}
\newacronym{mmtc}{mMTC}{massive machine type communication}
\newacronym{multi-rtt}{Multi-RTT}{multi-round trip rime}
\newacronym{nas}{NAS}{non-access stratum}
\newacronym{ng}{NG}{New Generation}
\newacronym{nlos}{NLOS}{Non-line-of-sight}
\newacronym{nrppa}{NRPPa}{NR positioning protocol A}
\newacronym{ofdm}{OFDM}{orthogonal frequency-division multiplexing}
\newacronym{otdoa}{OTDOA}{observed time-difference of arrival}
\newacronym{prs}{PRS}{positioning reference signal}
\newacronym{srs}{SRS}{sounding reference signal}
\newacronym{gdop}{GDOP}{geometric dilution of precision}
\newacronym{qos}{QoS}{Quality of service}
\newacronym{nr}{NR}{New Radio}
\newacronym{mmWave}{mmWave}{millimeter wave}

\newacronym{raim}{RAIM}{receiver autonomous integrity monitoring}
\newacronym{ran}{RAN}{Radio Access Network}
\newacronym{rat}{RAT}{Radio Access Technology}
\newacronym{rcs}{RCS}{radar cross-section}

\newacronym{rid}{remote-ID}{remote identification}
\newacronym{rtk}{RTK}{real-time kinematics}
\newacronym{rtt}{RTT}{Round Trip Time}
\newacronym{rx}{RX}{receiver}
\newacronym{sa}{SA}{Service \& System Aspects}
\newacronym{snr}{SNR}{signal-to-noise ratio}
\newacronym{tbs}{TBS}{Terrestrial Beacon System}
\newacronym{tdoa}{TDOA}{Time-Difference of Arrival}
\newacronym{toa}{TOA}{time of arrival}
\newacronym{tsg}{TSG}{technical specification group}
\newacronym{tx}{TX}{transmitter}
\newacronym{uas}{UAS}{unmanned aerial system}
\newacronym{uav}{UAV}{unmanned aerial vehicle}
\newacronym{ue}{UE}{user equipment}
\newacronym{ul}{UL}{uplink}
\newacronym{urllc}{uRLLC}{ultra reliable and low latency communication}
\newacronym{utdoa}{UTDOA}{uplink time difference of arrival}
\newacronym{utm}{UTM}{UAS traffic management system}
\newacronym{uwb}{UWB}{ultra-wideband}
\newacronym{v2x}{V2X}{vehicle-to-everything}
